\documentclass[twocolumn,pre,showpacs,showkeys]{revtex4-1}
\usepackage{graphicx}
\usepackage{soul}
\usepackage{color}
\usepackage{amsmath}
\usepackage{pgf}
\usepackage[greek,english]{babel}

\begin{document}

\title{Robust chimera states in SQUID metamaterials with local interactions} 

\author{J. Hizanidis, N. Lazarides, and G. P. Tsironis}

\affiliation{
Crete Center for Quantum Complexity and Nanotechnology, Department of Physics,
      University of Crete, P. O. Box 2208, 71003 Heraklion, Greece;  \\
Institute of Electronic Structure and Laser,
      Foundation for Research and Technology--Hellas, P.O. Box 1527, 71110 Heraklion,
      Greece;  \\		
National University of Science and Technology MISiS, Leninsky prosp. 4, Moscow, 
      119049, Russia
}

\date{\today}

\begin{abstract} 
We report on the emergence of robust multi-clustered chimera states in a dissipative-driven 
system of symmetrically and locally coupled identical SQUID oscillators. The ``snake-like'' 
resonance curve of the single SQUID (Superconducting QUantum Interference Device) is the 
key to the formation of the chimera states and is responsible for the extreme multistability 
exhibited by the coupled system that leads to attractor crowding at the geometrical resonance 
(inductive-capacitive) frequency. Until now, chimera states were mostly believed to exist for nonlocal coupling. 
Our findings provide theoretical evidence that nearest neighbor interactions are indeed 
capable of supporting such states in a wide parameter range. SQUID metamaterials are the 
subject of intense experimental investigations and we are highly confident that the complex 
dynamics demonstrated in this manuscript can be confirmed in the laboratory.
\end{abstract}


\maketitle

Since the first report on chimera states~\cite{KUR02a}, the number of works dedicated to 
this phenomenon of coexisting synchronous and desynchronous oscillatory behavior has grown 
immensely~(see \cite{panaggio:2015} and references within). The counterintuitive nature of 
chimeras inspired Abrams and Strogatz~\cite{ABR04} to name them after the mythological 
hybrid creature Chimera (Greek: \textgreek{Q\'imaira}) which has a lion's head, a goat's 
body and a snake's tail. The latest studies on chimera states focus on their stabilization 
and manipulation through various control techniques~\cite{SIE14,BIC15,ISE15,OME16} and their 
experimental verification~\cite{tinsley:2012,hagerstrom:2012,wickramasinghe:2013,martens:2013,
Rosin2014,schmidt:2014,Gambuzza2014,Kapitaniak2014}.

Chimera states have mostly been found for nonlocal coupling between the oscillators~\cite{OME13,
ZAK14,OME15}. This fact has given rise to a general notion that nonlocal coupling is an 
essential ingredient for their existence. However, recently, it has been demonstrated that 
chimeras can be achieved for global coupling too~\cite{schmidt:2014,SET14,Yeldesbay2014,BOE15}. 
The case of local coupling (i.e. nearest-neighbor interactions) has been studied less: In 
\cite{LAI15} chimera states were found in locally coupled networks, but the oscillators in 
the investigated systems were not completely identical. Very recently, the emergence of 
single- and double-headed (i.~e. with one and two (in)coherent regions, respectively) chimera states 
in neural oscillator networks with local coupling
has been reported~\cite{BER16}. That system, however, is known to exhibit high metastability, 
which renders the chimera state non-stationary when tracked in long time intervals~\cite{HIZ16}.
Here we demonstrate numerically the emergence of multi-clustered robust chimera states in 
SQUID metamaterials described in the local 
coupling approximation, in a relevant parameter region which has been determined experimentally 
in~\cite{Trepanier2013,Zhang2015}.

Superconducting metamaterials comprising SQUIDs have been realized in both one and two 
dimensions \cite{Trepanier2013,Zhang2015,Butz2013,Jung2014a,Jung2014b,Ustinov2015} and 
possess extraordinary properties such as negative magnetic permeability, dynamic multistability, 
broadband tunability, switching between different magnetic permeability states, as well as 
a unique form of transparency whose development can be manipulated through multiple parametric 
dependences. Some of these observed properties had been theoretically predicted both for the 
quantum \cite{Du2006} and the classical regime \cite{Lazarides2007,Lazarides2013}.
SQUID metamaterials are richly nonlinear effective media modeled by discrete phenomenological 
equations of coupled individual SQUID oscillators, which introduce qualitatively new 
macroscopic quantum effects into both the metamaterials and the coupled oscillator networks 
communities, i.e., magnetic flux quantization and the Josephson effect \cite{Josephson1962}.

\begin{figure}
\includegraphics[width=0.9\linewidth]{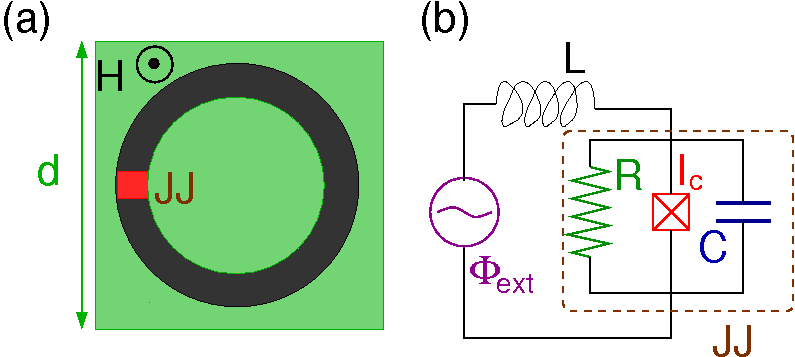}
\includegraphics[width=.9\linewidth]{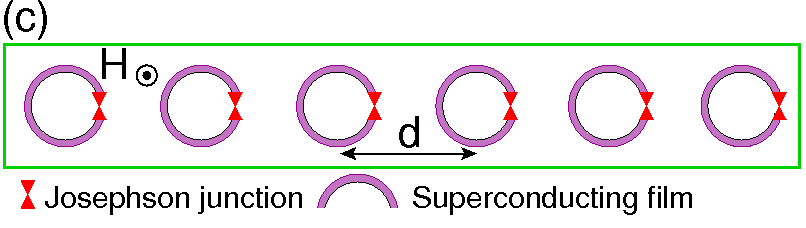}
\caption{(Color online)
Schematic of a SQUID in a magnetic field ${\bf H}(t)$ (a), and equivalent electrical 
circuit (b). The real Josephson junction is represented by the circuit elements in the 
brown-dashed box.
(c) Schematic of a one-dimensional SQUID metamaterial.
\label{fig1}
}
\end{figure}

A SQUID consists of a superconducting ring interrupted by a Josephson junction (JJ) as
shown schematically along with its electrical equivalent circuit in Figs. \ref{fig1}(a)
and \ref{fig1}(b), respectively; it is a highly nonlinear oscillator with a resonant response
to an applied alternating (ac) magnetic field. When a periodic arrangement of $N$ identical 
SQUIDs is driven by a spatially uniform, ac field [Fig. \ref{fig1}(c)], its elements are 
coupled together through magnetic dipole-dipole forces which decay as the inverse cube of 
the distance. In the following, it is considered that each SQUID in the array is coupled 
only to its nearest neighbors, neglecting further-neighbor interactions. Then, the magnetic 
flux $\Phi_n$ threading the loop of the $n$th SQUID is
\begin{eqnarray}
\label{01}
  \Phi_n =\Phi_{ext} +L\, I_n +M\, ( I_{n-1} +I_{n+1} ) ,
\end{eqnarray}
where $\Phi_{ext}$ is the external flux to each SQUID, $L$ is the self-inductance of the
individual SQUID, $M$ is the mutual inductance between neighboring SQUIDs, and
\begin{eqnarray}
\label{02}
    I_n =-C\frac{d^2\Phi_n}{dt^2} -\frac{1}{R} \frac{d\Phi_n}{dt} 
                                  -I_c\, \sin\left(2\pi\frac{\Phi_n}{\Phi_0}\right), 
\end{eqnarray}
is the current in the $n$th SQUID as provided by the resistively and capacitively shunted 
junction (RCSJ) model of the JJ \cite{Likharev1986}, and $\Phi_0$ is the flux quantum. 
Within the RCSJ framework, $R$, $C$, and $I_c$ are the resistance, capacitance, and 
critical current of the JJ, respectively. Combining Eqs. (\ref{01}) and (\ref{02}),
while neglecting all terms proportional to higher than the first power of the 
dimensionless coupling coefficient $\lambda=M/L$ \cite{Lazarides2013}, gives
\begin{eqnarray}
\label{05}
   \ddot{\phi}_n +\gamma \dot{\phi}_n +\phi_n +\beta \sin\left( 2\pi \phi_n \right) =
   \lambda ( \phi_{n-1} +\phi_{n+1} ) 
   \nonumber \\
   +(1 -2\lambda) \phi_{ac} \cos(\Omega \tau) , 
\end{eqnarray}
in which a sinusoidal external flux is considered. The flux through the $n$th SQUID loop 
$\phi_n$ and the amplitude of the external flux $\phi_{ac}$ are normalized to $\Phi_0$,
the driving frequency $\Omega$ and the temporal variable $\tau$ (the overdots denote 
derivation with respect to $\tau$) are normalized to the geometrical (inductive-capacitive)
resonance frequency of the SQUID $\omega_{LC} =1 / \sqrt{L C}$ and its inverse 
$\omega_{LC}^{-1}$, and $\beta=\frac{I_c L}{\Phi_0} =\frac{\beta_L}{2\pi}$, 
$\gamma=\frac{1}{R} \sqrt{ \frac{L}{C} }$ is the SQUID parameter and loss coefficient, 
respectively.

The corresponding equation for a single SQUID is obtained from Eq.~(\ref{05}) by setting 
$\lambda=0$ and $\phi_n =\phi$. Then, by linearization of that equation and by neglecting 
dissipation and forcing, the SQUID resonance frequency can be obtained as 
$\Omega_{SQ} =\sqrt{1 +\beta_L}$ in units of $\omega_{LC}$. 
The single SQUID equation for a certain range of parameters exhibits a ``snake-like'' 
resonance curve in which multiple stable and unstable periodic orbits coexist and vanish 
through saddle-node bifurcations of limit cycles. This dynamical behavior bears a big 
resemblance to the snaking bifurcation curves of localized structures reported in the 
Swift-Hohenberg equation \cite{Kozyreff2006}.

\begin{figure}
\includegraphics[width=\columnwidth]{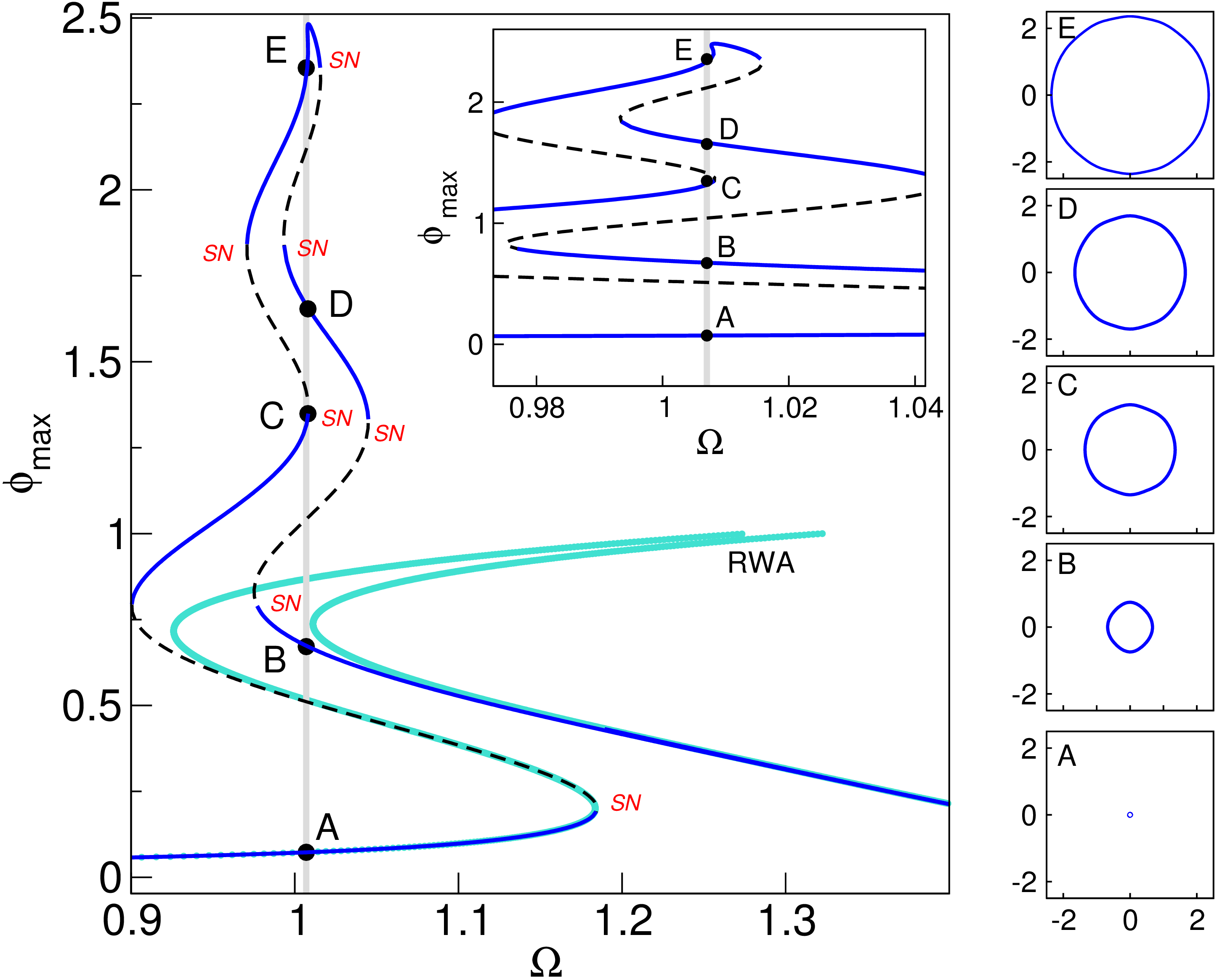}
\caption{(Color online)
The ``snake-like'' resonance curve of a single SQUID for $T=6.24$ ($\Omega\simeq 1.007$), 
$\beta_L = 0.86$, $\gamma=0.024$, and external ac flux $\phi_{ac}=0.06$.
Solid blue and dashed lines correspond to branches of stable and unstable periodic solutions, 
respectively. Saddle-node bifurcations points are denoted as ``SN''. Thick gray line 
corresponds to $\Omega=1.007$ and turquoise lines are obtained from Eq. (\ref{07}). 
Inset: Enlargement around the maximum multistability frequency. Phase portraits on the right 
show the corresponding periodic orbits of points $A-D$ on the resonance curve.
\label{fig2}}
\end{figure}

The parameters which are responsible for the SQUID multistability are the loss coefficient 
$\gamma$ and the external ac flux $\phi_{ac}$. As $\gamma$ decreases and $\phi_{ac}$ increases, 
the snaking curve becomes more winding, achieving higher flux values and adding, thus, to 
the multistability. A typical such curve is shown in the left panel of Fig.~\ref{fig2} in 
which the amplitude of the flux variable $\phi_{\text{max}}$ is plotted against the driving 
frequency $\Omega$. The blue solid lines correspond to the branches of stable periodic 
solutions while the dashed lines mark the unstable orbits. At each turning point where 
stable and unstable branches merge, a saddle node (``SN'') bifurcation of limit cycles 
takes place. The inset figure shows a blow-up around $\Omega=1$ where the multistability is 
more prominent. This is illustrated by the intersections of the gray line with the snaking 
curve marked by the letters $A-D$. For this value of the driving frequency, we can 
distinguish $K=5$ coexisting periodic states of increasing amplitude; the corresponding
orbits are shown in the phase portraits on the right.

An approximate resonance curve can be actually obtained from the single-SQUID equation
using a truncated series expansion for $\sin(2\pi \phi)$ with a  trial solution 
$\phi=\phi_m (\tau)\, \cos[\Omega \tau +\theta (\tau)]$, where $\phi_m (\tau)$ and 
$\theta (\tau)$ are the slowly varying amplitude and phase, respectively. Then, by applying 
the rotating wave approximation (RWA) in which only the terms at the fundamental frequency 
$\Omega$ are retained, neglecting terms $\propto \ddot{\phi_m}$, $\ddot{\theta}$, 
$\dot{\theta}^2$, $\dot{\phi_m}$, $\dot{\theta}$, etc., and seeking for steady state solutions
of the resulting algebraic system for $\phi_m (\tau)$ and $\theta (\tau)$, we get
\begin{equation}
\label{07}
   \Omega^2 =\Omega_{SQ}^2 -\beta_L \phi_m^2
              \{ a_1 -\phi_m^2 [a_2 -\phi_m^2 ( a_3 -a_4 \phi_m^2)]\} 
             \pm \frac{\phi_{ac}}{\phi_m} , 
\end{equation}
where $a_1=\pi^2/2$, $a_2=\pi^4/12$, $a_3=\pi^6/144$, $a_4=\pi^8/2880$, which implicitly 
provides the sought $\phi_m(\Omega)$ relation, in which the first four terms in the series 
expansion are kept. The curves obtained from the earlier equation are shown in Fig.~\ref{fig2} 
in turquoise color, and reproduce the resonance curve up to $\phi_m \sim 0.6$ that includes 
the first saddle-node bifurcation.

\begin{figure}
\includegraphics[width=0.5\columnwidth]{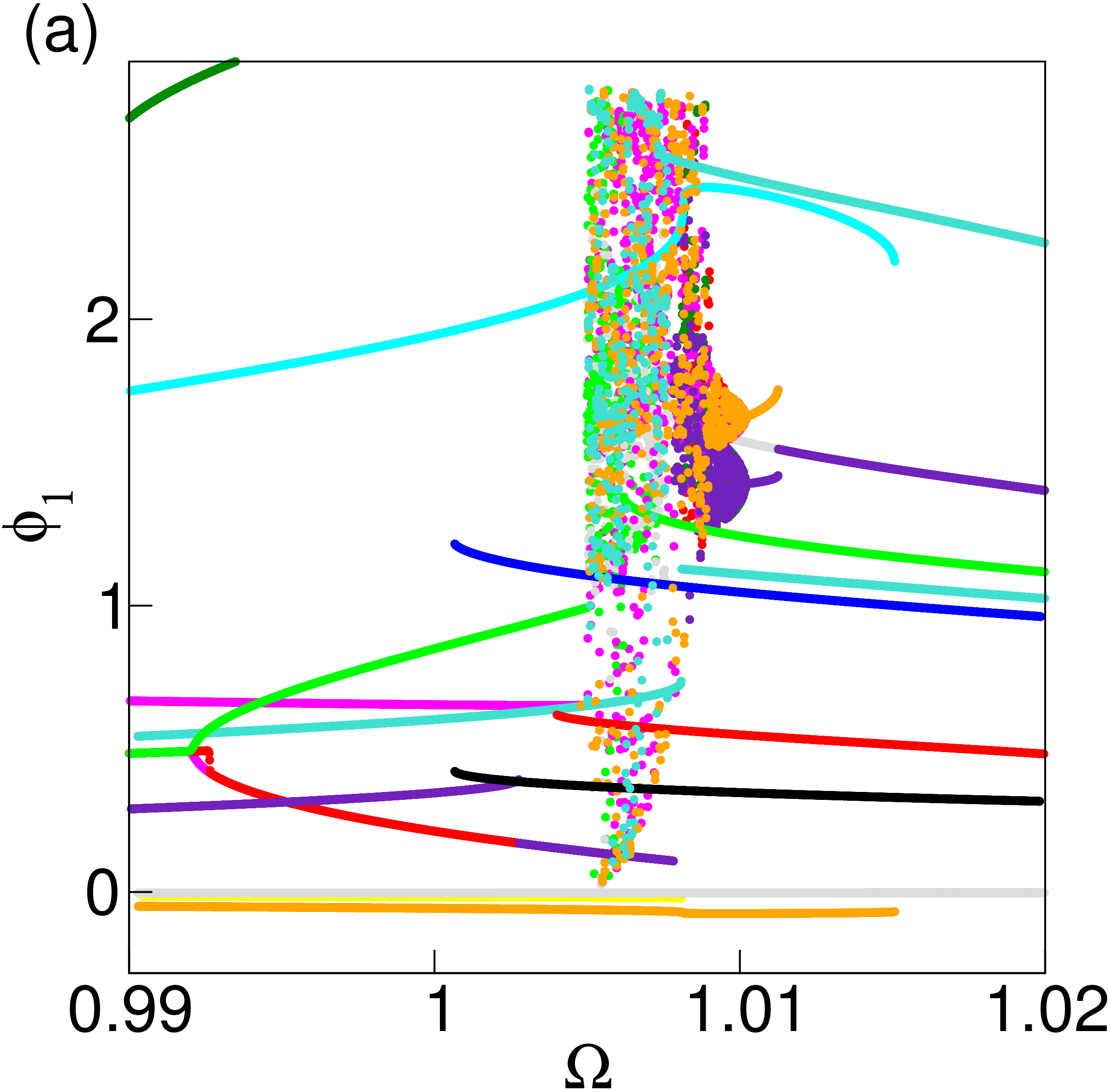} 
\includegraphics[width=0.46\columnwidth]{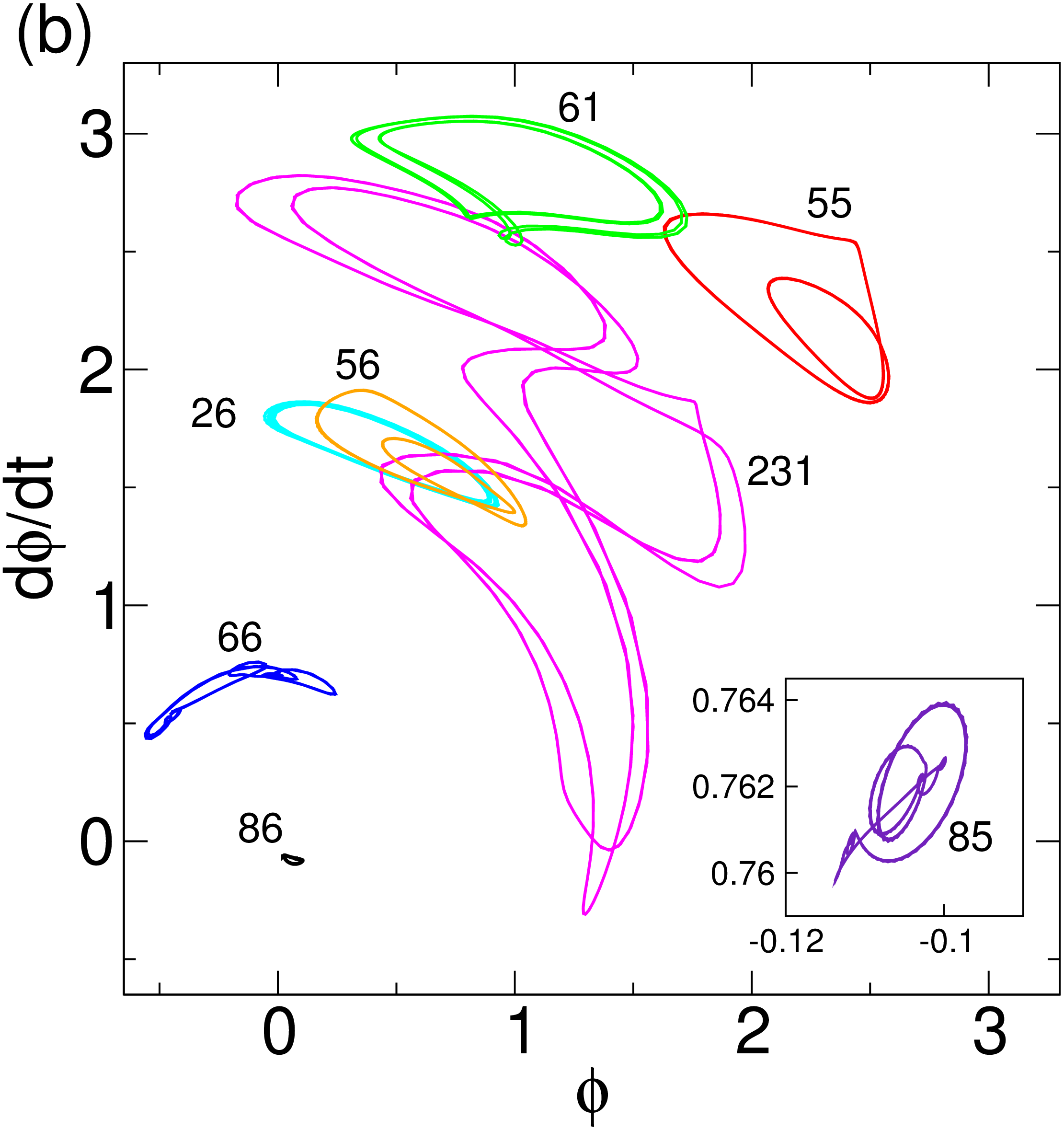} 
\caption{(Color online) 
(a) Solution branches for two coupled SQUIDs for an $\Omega$ scan around the maximum multistability point. At least 
    ten (10) SQUID states are visible at this frequency. 
(b) $N=256$ coupled SQUIDs: Stroboscopic maps of some individual oscillators. Numbers denote the index of the respective SQUID 
    oscillator. Coupling strength $\lambda=-0.025$ in both figures. All other parameters as 
    in Fig.~\ref{fig2}.
\label{fig3}
}
\end{figure}

In a metamaterial of $N$ weakly coupled SQUIDs there is a multiplicity of possible collective 
states that the system can reach, the number of which is of the order of $K^N$ or higher. 
The complexity in analyzing the behavior of such a system becomes clear by coupling together 
just two SQUIDs; the corresponding resonance curve maintains its ``snake-like'' form but 
with a thicker contour due to the additional (un)stable branches that are created (not 
shown here). Apart from the new periodic solutions, a number of coexisting chaotic attractors
are also to be found. Figure~\ref{fig3}(a) shows a scan in $\Omega$ for values around the maximum multistability 
point, for two coupled SQUIDs. The different colors correspond 
to solution branches for different initial conditions and it is clear that around $\Omega=1.007$ the magnetic flux
exhibits chaotic behavior. For $N=256$ coupled SQUIDs the complexity of the dynamics
is even higher: Figure~\ref{fig3}(b) shows the stroboscopic maps corresponding to individual oscillators of a 
single configuration of the full system.    
The numbers next to the orbits denote the oscillator indices. This huge multiplicity of attractors is known as 
{\em attractor crowding} and has been observed before in coupled nonlinear oscillator 
systems \cite{Wiesenfeld1989}.

\begin{figure*}
\includegraphics[width=.44\textwidth]{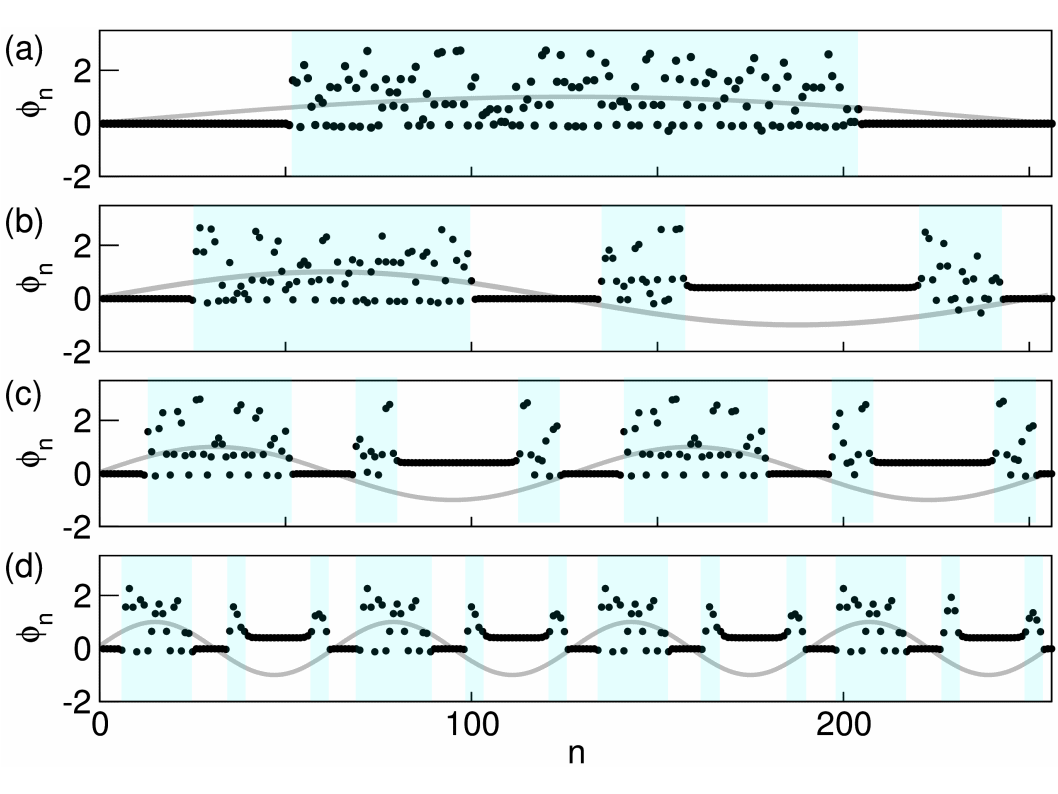}  
\includegraphics[width=.44\textwidth]{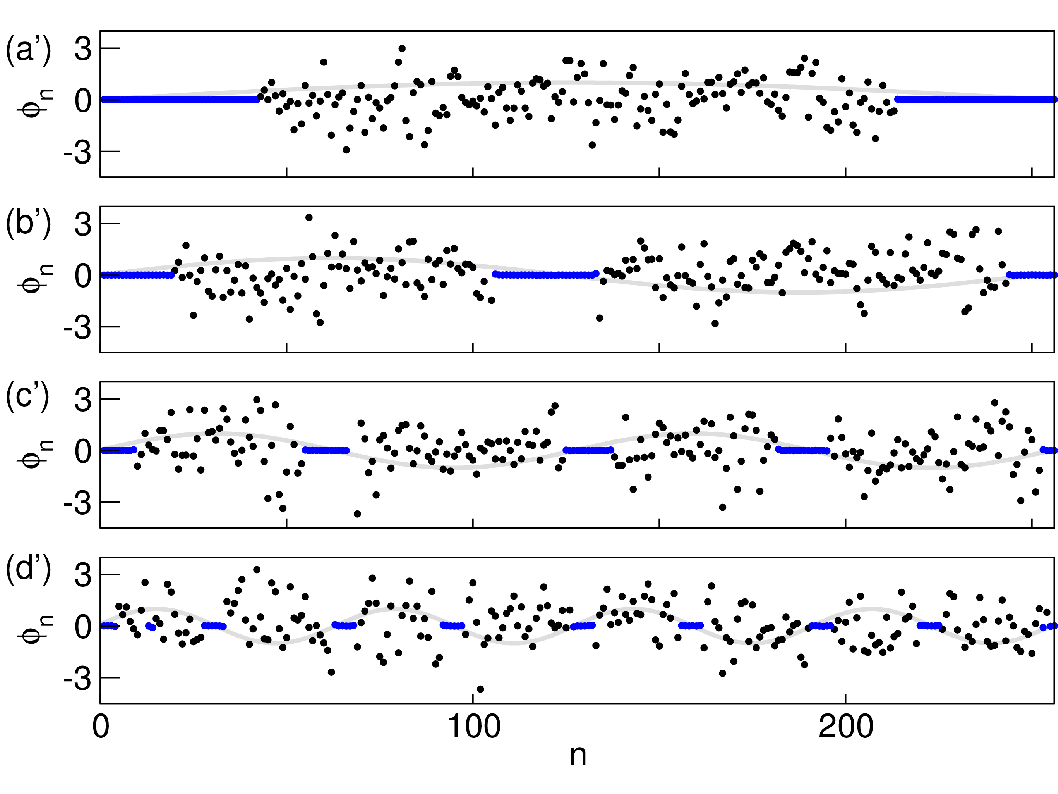} 
\caption{(Color online)
Snapshots of the magnetic fluxes $\phi_n$ at time $T=5000$ for two different values of the 
loss coefficient: $\gamma=0.024$ in (a)-(d) and $\gamma=0.0024$ in (a')-(d'). Grey solid lines
mark the initial magnetic flux distribution used in the simulations. Colored areas in the left panel
mark the incoherent clusters while blue solid lines
in the right panel emphasize the coherent clusters of the chimera states. All other 
parameters as in Fig.~\ref{fig2}. See Supplemental Material for animations related to
Figs.~\ref{fig4}(b) \cite{SM1} and (b') \cite{SM2}. 
\label{fig4}
}
\end{figure*}

Equations (\ref{05}) are integrated numerically in time with a fourth-order Runge-Kutta 
algorithm with constant time-step and periodic boundary conditions, i.~e., 
$\phi_n(\tau)=\phi_{N+n}(\tau)$ for all $n$. The parameters used in the simulations are 
close to the design parameters of the SQUID meta-atoms that make two-dimensional SQUID 
metamaterials \cite{Zhang2015}, i.e., $L=60~pH$, $C=0.42~pF$, $I_c=4.7~\mu A$, and subgap 
resistance $R=500~\Omega$, which give $\beta_L =0.86$ and $\gamma=0.024$ according to their
definitions, while the value of the coupling coefficient between neighboring SQUIDs is 
$\lambda=-0.025$. The amplitude of the ac field is selected to be $\phi_{ac}=0.06$, 
within the experimentally accessible range $0.001 -0.1$ \cite{Trepanier2013}.
The selected values of $\gamma$ and $\phi_{ac}$ bring the SQUID metamaterial in the strongly
nonlinear regime.

Figure~\ref{fig4} shows time-snapshots of the magnetic fluxes $\phi_n$ for different initial 
conditions and for two different values of the loss coefficient $\gamma$. The left panel is 
for $\gamma=0.024$ which is the value corresponding to the resonance curve of Fig.~\ref{fig2}.
The initial ``sine wave'' magnetic flux distribution for each simulation is shown by the gray 
solid line. The SQUIDs that are prepared at lower values form the coherent clusters of the 
chimera state, while those that are initially set at higher magnetic flux values oscillate 
incoherently. Moreover, as the ``wave-length'' of the initial magnetic flux distribution 
increases, so does the chimera state multiplicity (number of (in)coherent regions highlighted by the colored areas).
Similar behavior is observed for lower values of the loss coefficient ($\gamma=0.0024$) as 
shown in the right panel of Fig.\ref{fig4}. Here, the incoherent clusters are better 
illustrated since they are approximately of equal size and do not contain oscillators that 
``escape'' from the incoherent cluster abiding around low magnetic flux values, something 
which is visible in the left panel. Furthermore, the coherent clusters (emphasized by the 
blue solid lines) are fixed around $\phi=0$, unlike in the left panel where additional clusters 
located at slightly higher values also form. Here we must recall that for low values of 
$\gamma$ (right panel) the winding of the ``snake-like'' resonance curve increases 
significantly creating, thus, new branches of stable (and equally unstable) periodic solutions.
These branches are larger in number and smaller in size compared to those of higher $\gamma$ 
values (left panel). The lower amplitude branches which are the longer ones attract the 
SQUIDs that eventually form the coherent clusters. The other oscillators have a plethora 
of higher states to choose from and, therefore, create a more chaotic incoherent cluster than 
in the case of higher $\gamma$ values. 

\begin{figure}[!h]
\includegraphics[width=0.47\columnwidth]{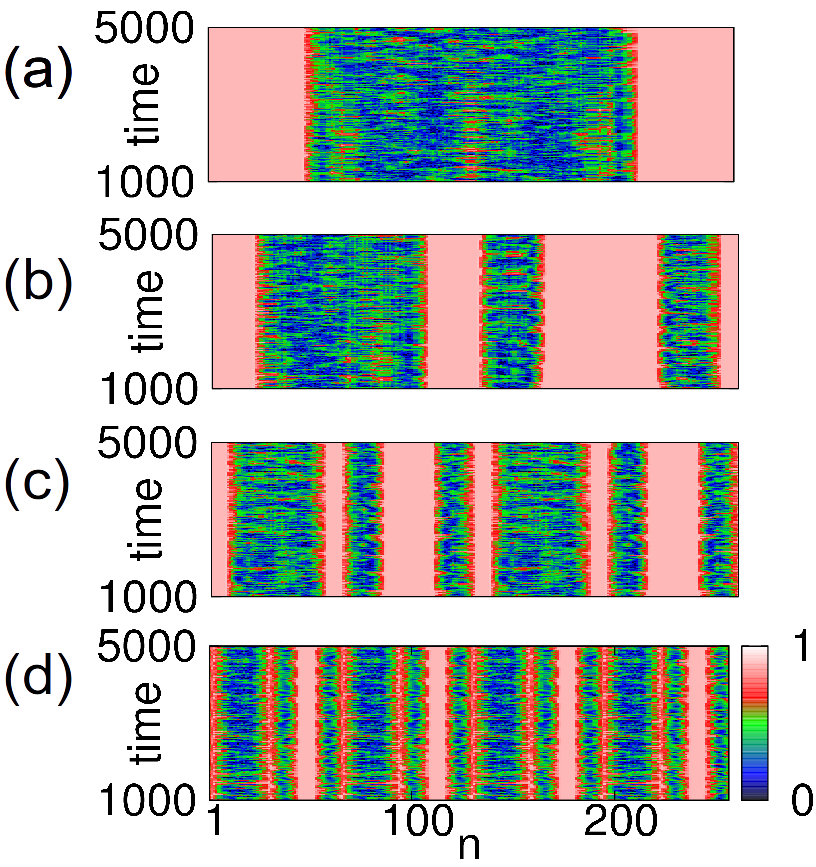} 
\includegraphics[width=0.51\columnwidth]{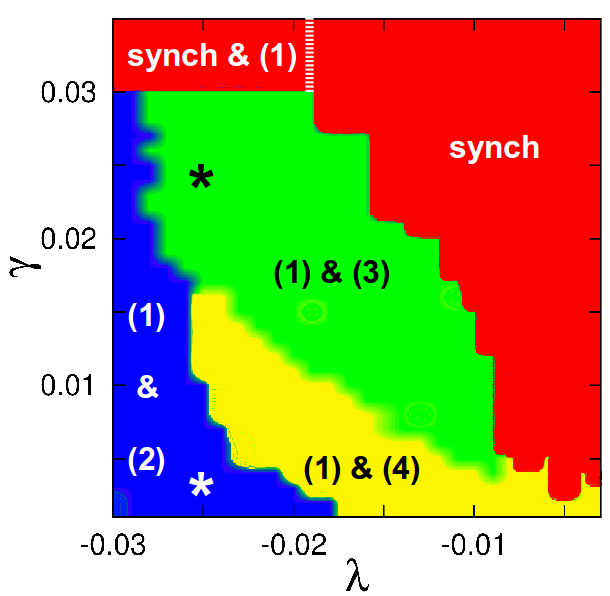}
\caption{(Color online)
Left panel: Space-time plots for the magnitude of the local order parameter $|Z_n|$ of the 
chimera states corresponding to Figs.~\ref{fig4}(a)-(d).
Right panel: Map of dynamic regimes in the $(\gamma,\lambda)$ parameter space
for the initial conditions of Figs.~\ref{fig4}(a) and (b).
Numbers in brackets denote the multiplicity of the chimera state while ``synch'' stands for 
(route to) synchronization. All other parameters as in Fig.~\ref{fig2}.
\label{fig5}
}
\end{figure}

The observed chimera states can be quantified through the Kuramoto local order parameter 
\cite{HIZ16a} which is a measure for local synchronization: 
\begin{equation}
 Z_n=\left | \frac{1}{2\delta} \sum_{|j-n| \le \delta} e^{i\phi_j} \right |, \quad n=1,\dots, N.
\end{equation}
We use a spatial average with a window size of $\delta=5$ elements. A $Z_n$ value close to 
unity indicates that the $n$th SQUID belongs to the coherent cluster of the chimera state, 
while $Z_n$ is closer to $0$ in the incoherent parts. In the left panel of Fig.~\ref{fig5} 
the space-time plots of the local order parameter corresponding to the chimera states of 
Fig.~\ref{fig4}(a)-(d) are shown. The number of (in)coherent regions increases according to 
the initial conditions and the size and location of the clusters is constant in time. Previous 
works on SQUID metamaterials showed that for nonlocal coupling single and double-headed 
chimera states coexist with solitary states~\cite{Jaros2015} and metastable states of drifting 
(in)coherence \cite{LAZ15,HIZ16a}. Note that in these studies the focus was on a different 
dynamical area with the external driving frequency lying outside the multistability regime.
For a suitable choice of $\Omega$, stable chimera states can be achieved for nonlocal
coupling also. However, they exist only for low coupling strengths $\lambda$; the threshold 
value of the coupling strength in the case of local coupling is much higher. Local coupling 
is therefore crucial for the emergence of \emph{robust} chimera states, both in structure and 
in lifetime, for a wide range of parameters.

Previously we stressed the importance of multistability and the impact of the loss coefficient 
$\gamma$ in the formation of chimeras in our system. In addition to that, it is important to 
note the role of the network topology which is defined through the local nature of interactions 
and the coupling strength $\lambda$. As already shown in Fig.~\ref{fig4}, our system exhibits 
a variety of coexisting multi-clustered chimera states. A systematic study in the 
$(\lambda, \gamma)$ parameter space is depicted in the right panel of Fig.~\ref{fig5}, where 
the observed patterns for two different sets of initial conditions (namely those of 
Fig.~\ref{fig4}(a,a') and (b,b')) are mapped out. The numbers in the brackets correspond to 
the multiplicity of the respective chimeras and ``synch'' denotes the synchronized states.
The black and white asterisk mark the $(\lambda, \gamma)$ values used in the left and right 
panel of Fig.~\ref{fig4}, respectively. We see that for low values of the loss coefficient
and for low and medium (in absolute value) coupling strengths, single- and 
four-headed chimera states coexist (yellow area). As $\gamma$ 
increases, the effect of multistability diminishes and the system enters the synchronized state (red area) either directly 
or through a region where three-headed chimeras coexist with single-headed ones (green area).
For stronger couplings (blue area), double-headed 
chimeras coexist with single chimeras. The latter persist also for high $\gamma$ values where the synchronized state is achieved.   
For initial conditions with a larger modification in space (like in 
Fig.~\ref{fig4}(c,c') and (d,d')) chimera states with higher multiplicity can be found, 
but the mechanism towards synchronization is the same: the fully coherent state is reached through 
the appearance of solitary states~\cite{Jaros2015,HIZ16a}.

In conclusion, the model equations for a SQUID metamaterial truncated to only nearest-neighbor 
coupling were integrated in time with properly chosen initial conditions that allow the system 
to reach chimera states. These novel states emerge due to the extreme multistability that leads 
to attractor crowding at the geometrical resonance frequency. Typical chimera states are 
presented and characterized with respect to their local synchronization level. A systematic 
study in the relevant parameter space reveals the coexistence of multi-headed chimeras and the 
oscillators for a metamaterial in a chimera state elucidate a number of different trajectories, 
some of which are chaotic. Since chimera states have been intimately connected with nonlocal
coupling, the present results point towards the need to revise the general consensus
on the essential conditions for their existence.

\subsection*{Acknowledgement}
This work was partially supported by
the European Union Seventh Framework Programme (FP7-REGPOT-2012-2013-1) under grant
agreement n$^o$ 316165,
and the Ministry of Education and Science of the Russian Federation in the framework of the 
Increase Competitiveness Program of NUST  ``MISiS'' (No. K2-2015-007).
J. H. would like to thank Thomas Isele and Yuri Maistrenko for valuable discussions.


\begin{thebibliography}{10}

\bibitem{KUR02a} 
  Y. Kuramoto and D. Battogtokh, 
  Nonlinear Phenom. Complex Syst. \textbf{5}, 380 (2002).

\bibitem{panaggio:2015}
  M.~J. Pannagio and D.~Abrams,  
  Nonlinearity \textbf{28}, R67 (2015).
  
  
\bibitem{ABR04} 
  D.~M. Abrams and S. H. Strogatz, Phys. Rev. Lett. \textbf{93}, 174102 (2004).
  
\bibitem{SIE14} 
  J. Sieber, O. E. Omel'chenko, and M. Wolfrum, 
  Phys. Rev. Lett. \textbf{112}, 054102 (2014).
  
\bibitem{BIC15} 
  C. Bick and E.~A. Martens, New J. Phys. \textbf{17}, 033030 (2015).

\bibitem{OME16}
  I. Omelchenko, O.~E. Omel'chenko, A. Zakharova, M. Wolfrum, and E. Sch{\"o}ll,
  Phys. Rev. Lett. \textbf{116}, 114101 (2016).
  
\bibitem{ISE15}
  T. Isele, J. Hizanidis, A. Provata, and P. H\"ovel,
  Phys. Rev. E \textbf{93}, 022217 (2016).
  
\bibitem{tinsley:2012} 
  M. R. Tinsley, S. Nkomo, and K. Showalter, 
  Nature Phys. \textbf{ 8}, 662 (2012).

\bibitem{hagerstrom:2012}
  A.~M.Hagerstrom, T.~E. Murphy, R.~Roy, P.~H{\"o}vel, I.~Omelchenko, and E.~Sch{\"o}ll, 
  Nature Phys. \textbf{8}, 658 (2012).
  
\bibitem{wickramasinghe:2013}
  M.~Wickramasinghe and I.~Z. Kiss, 
  PLoS ONE \textbf{8}, e80586 (2013).
  
\bibitem{martens:2013}
  E.~A. Martens, S.~Thutupalli, A.~Fourri{\`e}re, and O.~Hallatschek, 
  Proc. Nat. Acad. Sciences \textbf{110}, 10563 (2013).
  
\bibitem{Rosin2014}
D.~P. Rosin, D. Rontani, N.~D. Haynes,~E. Sch\"oll, and D.~J. Gauthier, 
  Phys. Rev. E \textbf{90}, 030902 (2014). 


\bibitem{schmidt:2014}
  L. Schmidt, K. Sch\"onleber, K. Krischer, and V. Garc\'ıa-Morales,
  Chaos \textbf{24}, 013102 (2014).
  
\bibitem{Gambuzza2014}
  L. V. Gambuzza, A. Buscarino, S. Chessari, L. Fortuna, R. Meucci, and M. Frasca,
  Phys. Rev. E \textbf{90}, 032905 (2014).

\bibitem{Kapitaniak2014}
  T. Kapitaniak, P. Kuzma, J. Wojewoda, K. Czolczynski, and Y. Maistrenko,
  Sci. Rep. \textbf{4}, 6379 (2014).  

\bibitem{OME13} 
  I. Omelchenko, O. E. Omel'chenko, P. H{\"o}vel, and E. Sch{\"o}ll,
  Phys. Rev. Lett. \textbf{110}, 224101 (2013).
  
\bibitem{ZAK14} 
  A. Zakharova, M. Kapeller, and E. Sch\"oll, 
  Phys. Rev. Lett. \textbf{112}, 154101 (2014).
  
\bibitem{OME15}
  I.~Omelchenko, A.~Provata, J.~Hizanidis, E.~Sch{\"o}ll, and P.~H\"ovel,
  Phys. Rev. E {\bf 91}, 022917 (2015).

\bibitem{SET14}
  G. C. Sethia and A. Sen,
  Phys. Rev. Lett. \textbf{112}, 144101 (2014).

\bibitem{Yeldesbay2014} 
  A. Yeldesbay, A. Pikovsky, and M. Rosenblum, 
  Phys. Rev. Lett. \textbf{112}, 144103 (2014).
  
\bibitem{BOE15} 
  F. B\"ohm, A. Zakharova, E. Sch\"oll, and K. L\"udge,
  Phys. Rev. E \textbf{91}, 040901(R) (2015). 
  
\bibitem{LAI15} 
  C. Laing, 
  Phys. Rev. E \textbf{92}, 050904(R) (2015).
  
\bibitem{BER16}
  B. K. Bera, D. Ghosh, and M. Lakshmanan
  Phys. Rev. E \textbf{93}, 012205 (2016).

\bibitem{HIZ16}
  J. Hizanidis, N. E. Kouvaris, G. Zamora-L\'opez, A. D\'iaz-Guilera, and C. G. Antonopoulos,
  Sci. Rep. \textbf{6}, 19845 (2016).

\bibitem{Trepanier2013} 
  M. Trepanier, D. Zhang, O. Mukhanov, and S.~M. Anlage, 
  Phys. Rev. X \textbf{3}, 041029 (2013).
  
  \bibitem{Zhang2015}
  D. Zhang, M. Trepanier, O. Mukhanov, and S. M. Anlage,
  Phys. Rev. X \textbf{5}, 041045 (2015).
  
\bibitem{Butz2013}
   S. Butz, P. Jung, L.~V. Filippenko, V.~P. Koshelets, and A.~V. Ustinov,
   Opt. Express \textbf{21}, 22540 (2013).

\bibitem{Jung2014a}
   P. Jung, S. Butz, M. Marthaler, M. V. Fistul, J. Lepp\"akangas, V. P. Koshelets,
   and A. V. Ustinov,
   Nat. Commun. \textbf{5}, 3730 (2014).

\bibitem{Jung2014b}
   P. Jung, A. V. Ustinov, and S. M. Anlage,
   Supercond. Sci. Technol. \textbf{27}, 073001 (2014).

\bibitem{Ustinov2015}
A. V. Ustinov, IEEE Trans. Terahertz Sci. Technol. \textbf{5}, 22 (2015).

\bibitem{Du2006}
   C. Du, H. Chen, and S. Li, Phys. Rev. B \textbf{74}, 113105 (2006).

\bibitem{Lazarides2007}
   N. Lazarides and G. P. Tsironis, Appl. Phys. Lett. \textbf{90}, 163501 (2007).

\bibitem{Lazarides2013} 
  N. Lazarides and G. P. Tsironis, 
  Supercond. Sci. Technol. \textbf{26}, 084006 (2013).
  
\bibitem{Josephson1962} 
  B. Josephson, Phys. Lett. A \textbf{1}, 251 (1962).

\bibitem{Likharev1986}
  K.~K. Likharev
  \emph{Dynamics of Josephson Junctions and Circuits},
  Gordon and Breach, Philadelphia (1986).
  
\bibitem{Kozyreff2006}
  G. Kozyreff and S.~J. Chapman,
  Phys. Rev. Lett. \textbf{97}, 044502 (2006).
  
\bibitem{Wiesenfeld1989}
  K. Wiesenfeld and P. Hadley,
  Phys. Rev. Lett. \textbf{62}, 1335 (1989).

\bibitem{SM1}
Supplemental video SM1
\bibitem{SM2}
Supplemental video SM2
  
\bibitem{HIZ16a}
  J. Hizanidis, N. Lazarides, G. Neofotistos, and G. P. Tsironis,
  Eur.~Phys.~J. Special Topics, Springer 225 (5) (2016) (in print). 

\bibitem{Jaros2015} 
  P. Jaros, Y. Maistrenko, and T. Kapitaniak,
  Phys. Rev. E \textbf{91}, 022907 (2015).

\bibitem{LAZ15} 
  N. Lazarides, G. Neofotistos, and G.~P. Tsironis,
  Phys. Rev. B \textbf{91}, 054303 (2015).

  
\end{thebibliography}
\end{document}